\documentclass[conference]{IEEEtran}
\usepackage{ifpdf}
\usepackage[nospace]{cite}
\usepackage{graphicx}
\usepackage{color}
\usepackage{bm}
\usepackage[cmex10]{amsmath}
\usepackage{amssymb}
\usepackage{array}
\usepackage{syntonly}
\usepackage{amsthm}
\usepackage{amsfonts}
\usepackage{epsfig}
\usepackage{latexsym}
\newcommand{\Define}{\stackrel{\triangle}{=}}

\usepackage{fixltx2e}
\usepackage{url}
\theoremstyle{definition} 
\theoremstyle{definition} 
\theoremstyle{definition} 
\begin{document}
\title{{\LARGE Sum Secrecy Rate in Full-Duplex Wiretap Channel with Imperfect 
CSI} }
\author{\IEEEauthorblockN{Sanjay Vishwakarma}
\IEEEauthorblockA{Dept. of ECE\\
Indian Institute of Science\\
Bangalore 560012, India\\
sanjay@ece.iisc.ernet.in}
\and
\IEEEauthorblockN{A. Chockalingam}
\IEEEauthorblockA{Dept. of ECE\\
Indian Institute of Science\\
Bangalore 560012, India\\
Email: achockal@ece.iisc.ernet.in}}
\vspace{-30mm}
\author{{\large Sanjay Vishwakarma and A. Chockalingam} \\
sanjay@ece.iisc.ernet.in, achockal@ece.iisc.ernet.in \\
{\normalsize Department of ECE, Indian Institute of Science,
Bangalore 560012, India}
}
\maketitle
\begin{abstract}
In this paper, we consider the achievable sum secrecy rate in full-duplex 
wiretap channel in the presence of an eavesdropper and imperfect channel 
state information (CSI). We assume that the users participating in 
full-duplex communication and the eavesdropper have single antenna each. 
The users have individual transmit power constraints. They also transmit 
jamming signals to improve the secrecy rates. We obtain the achievable 
perfect secrecy rate region by maximizing the sum secrecy rate. We also 
obtain the corresponding optimum powers of the message signals and the 
jamming signals. Numerical results that show the impact of imperfect CSI 
on the achievable secrecy rate region are presented.
\end{abstract}
{\em keywords:}
{\em {\footnotesize
Full-duplex, physical layer security, secrecy rate,
linear and semidefinite programming.
}} 
\IEEEpeerreviewmaketitle

\section{Introduction}
\label{sec1}
Transmitting messages with perfect secrecy using physical layer 
techniques was first studied in \cite{ic1} on a physically degraded 
discrete memoryless wiretap channel model. Later, this work was 
extended to more general broadcast channel in \cite{ic2} and Gaussian 
channel in \cite{ic3}, respectively. Wireless transmissions, being 
broadcast in nature, can be easily eavesdropped and hence require 
special attention to design modern secure wireless networks. Secrecy 
rate and capacity of point-to-point multi-antenna wiretap channels
have been reported in the literature by several authors, e.g., 
\cite{ic4, ic6, ic7, ic8, ic9, ic10}. In the above works, the 
transceiver operates in half-duplex mode, i.e., either it transmits 
or receives at any given time instant. On the other hand, full-duplex 
operation gives the advantage of simultaneous transmission and reception 
of messages \cite{fd_rice}. But loopback self-interference and imperfect 
channel state information (CSI) are limitations. Full-duplex communication 
without secrecy constraint has been investigated by many authors, e.g., 
\cite{ic11, ic12, ic13, ic14}. Full-duplex communication with secrecy 
constraint has been investigated in \cite{ic20, ic21, ic22}, where the 
achievable secrecy rate region of two way Gaussian and discrete memoryless 
wiretap channels have been characterized. In the above works, CSI in all 
the links are assumed to be perfect.

In this paper, we consider the achievable sum secrecy rate in full-duplex 
wiretap channel in the presence of an eavesdropper and imperfect CSI. The 
both users participating in full-duplex communication and the eavesdropper 
are assumed to have single antenna each. The CSI errors in all the links 
are assumed to be bounded in their respective absolute values. In addition 
to a message signal, each user transmits a jamming signal in order to 
improve the secrecy rates. The users operate under individual power 
constraints. For this scenario, we obtain the achievable perfect secrecy 
rate region by maximizing the sum secrecy rate. We also obtain the 
corresponding optimum powers of the message signals and jamming signals. 
Numerical results that illustrate the impact of imperfect CSI on the 
achievable secrecy rate region are presented.

The rest of the paper is organized as follows. The system model is
given In Sec. \ref{sec2}. Secrecy rate for perfect CSI is presented in
\ref{sec3}. Secrecy rate with imperfect CSI is studied in \ref{sec4}.
Results and discussions are presented in Section \ref{sec5}. Conclusions
are presented in Section \ref{sec6}. 

$\bf{Notations:}$ 
$\boldsymbol{A} \succeq \boldsymbol{0}$ implies that $\boldsymbol{A}$ is 
a positive semidefinite matrix. Transpose and complex conjugate transpose 
operations are denoted by $[.]^{T}$ and $[.]^{\ast}$, respectively.
$\lvert.\rvert$ denotes absolute value operation.

\section{System Model}
\label{sec2}
\begin{figure}
\center
\includegraphics[totalheight=6.5cm,width=8.5cm]{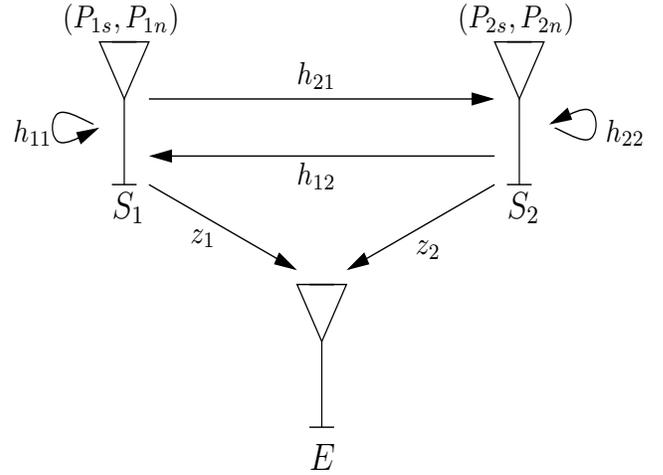}
\caption{System model for full-duplex communication.}
\label{fig1}
\end{figure}

We consider full-duplex communication between two users $S_{1}$ and $S_{2}$
in the presence of an eavesdropper $E_{}$. $S_{1}$, $S_{2}$ and $E_{}$ are 
assumed to have single antenna each. The complex channel gains on various 
links are as shown in Fig. \ref{fig1}.
$S_{1}$ and $S_{2}$ simultaneously transmit messages $W_{1}$ and $W_{2}$, 
respectively, in $n$ channel uses. $W_{1}$ and $W_{2}$ are independent and 
equiprobable over $\{1,2,\cdots,2^{nR_{1}}\}$ and $\{1,2,\cdots,2^{nR_{2}}\}$,
respectively. $R_{1}$ and $R_{2}$ are the information rates (bits per channel 
use) associated with $W_{1}$ and $W_{2}$, respectively, which need to be 
transmitted with perfect secrecy with respect to $E_{}$ \cite{ic22}. 
$S_{1}$ and $S_{2}$ map $W_{1}$ and $W_{2}$ to i.i.d. codewords 
$\{X^{1}_{i}\}^{n}_{i = 1}$ $\big(\sim \mathcal{CN}(0, P_{1s}) \big)$ 
and $\{X^{2}_{i}\}^{n}_{i = 1}$ $\big(\sim \mathcal{CN}(0, P_{2s}) \big)$, 
respectively, of length $n$. In order to degrade the eavesdropper channels 
and improve the secrecy rates, both $S_{1}$ and $S_{2}$ transmit i.i.d. 
jamming signals
$\{{N}^{1}_{i}\}^{n}_{i = 1}$ $\big(\sim \mathcal{CN}(0, P_{1n}) \big)$ and
$\{{N}^{2}_{i}\}^{n}_{i = 1}$ $\big(\sim \mathcal{CN}(0, P_{2n}) \big)$, 
respectively, of length $n$. $S_{1}$ and $S_{2}$ transmit the symbols 
$X^{1}_{i} + N^{1}_{i}$ and $X^{2}_{i} + N^{2}_{i}$, respectively, during
the $i$th channel use, $1 \leq i \leq n$. Hereafter, we will denote the 
symbols in $\{{X}^{1}_{i}\}^{n}_{i = 1}$, $\{{X}^{2}_{i}\}^{n}_{i = 1}$  
$\{{N}^{1}_{i}\}^{n}_{i = 1}$, $\{{N}^{2}_{i}\}^{n}_{i = 1}$ with 
${X}^{1}_{}, \ {X}^{2}_{}, \ {N}^{1}_{} $ and ${N}^{2}_{}$, respectively.
We also assume that all the channel gains remain static over the codeword 
transmit duration. Let $P_{1}$ and $P_{2}$ be the transmit power budget for 
$S_{1}$ and $S_{2}$, respectively. This implies that 
\begin{eqnarray}
[P_{1s} + P_{1n}, \ P_{2s} + P_{2n}]^{T} \ \leq \ [P_{1}, \ P_{2}]^{T}. 
\label{eqn52}
\end{eqnarray}
Let $y_{1}$, $y_{2}$ and $y_{E}$ denote the received signals at $S_{1}$, 
$S_{2}$ and $E_{}$, respectively.
We have 
\begin{eqnarray}
y_{1}   &=& h_{11}(X^{1} + N^{1}) +  h_{12}(X^{2} + N^{2}) + \eta_{1}, 
\label{eqn1} \\
y_{2}   &=& h_{21}(X^{1} + N^{1}) +  h_{22}(X^{2} + N^{2}) + \eta_{2}, 
\label{eqn2} \\
y_{E} &=& z_{1}(X^{1} + N^{1}) +  z_{2}(X^{2} + N^{2}) + \eta_{E},        
\label{eqn3}
\end{eqnarray}
where $\eta_{1}$, $\eta_{2}$ and $\eta_{E}$ are i.i.d. 
$(\sim \mathcal{CN}(0, N_{0}))$ receiver noise terms.

\section{Sum secrecy rate - perfect CSI}
\label{sec3}
In this section, we assume perfect CSI in all the links. Since $S_{1}$ 
knows the transmitted symbol $(X^{1} + N^{1})$, in order to detect $X^{2}$, 
$S_{1}$ subtracts ${h}^{}_{11}(X^{1} + N^{1})$ from the received signal 
$y_{1}$, i.e.,
\begin{eqnarray}
y^{'}_{1} = y_{1} - h_{11}(X^{1} + N^{1}) = h_{12}(X^{2} + N^{2}) + \eta_{1}. 
\label{eqn8} 
\end{eqnarray}
Similarly, since $S_{2}$ knows the transmitted symbol $(X^{2} + N^{2})$, 
to detect $X^{1}$, $S_{2}$ subtracts ${h}^{}_{22}(X^{2} + N^{2})$ from 
the received signal $y_{2}$, i.e., 
\begin{eqnarray}
y^{'}_{2} = y_{2} - h_{22}(X^{2} + N^{2}) = h_{21}(X^{1} + N^{1}) + \eta_{2}. 
\label{eqn9} 
\end{eqnarray}
Using (\ref{eqn8}) and (\ref{eqn9}), we get the following information rates 
for $X^{1}$ and $X^{2}$, respectively:
\begin{eqnarray}
R^{'}_{1} \Define I\big(X^{1}; \ y^{'}_{2}\big) = \log_{2} \Big( 1 + \frac{{\lvert h_{21}\rvert}^{2}P_{1s}}{N_{0} + {\lvert h_{21}\rvert}^{2}P_{1n}}\Big),
\label{eqn54} \\
R^{'}_{2} \Define I\big(X^{2}; \ y^{'}_{1}\big) = \log_{2} \Big( 1 + \frac{{\lvert h_{12}\rvert}^{2}P_{2s}}{N_{0} + {\lvert h_{12}\rvert}^{2}P_{2n}}\Big).
\label{eqn53} 
\end{eqnarray}
Using (\ref{eqn3}), we get the following information leakage rate at $E$:
\begin{eqnarray}
R^{'}_{E} & \Define & I\big(X^{1}, X^{2}; \   y^{}_{E}\big) \nonumber \\ 
& = & \log_{2} \Big( 1 + \frac{ {\lvert z_{1}\rvert}^{2}P_{1s} + {\lvert z_{2}\rvert}^{2}P_{2s} }{ N_{0} + {\lvert z_{1} \rvert}^{2}P_{1n} + {\lvert z_{2}\rvert}^{2}P_{2n} } \Big). 
\label{eqn55}
\end{eqnarray}
We denote the information capacities by $C^{}_{1}$, $C^{}_{2}$, and $C^{}_{E}$, respectively, as follows:
\begin{eqnarray}
C_{1} \  = \ \log_{2} \Big( 1 + \frac{{\lvert h_{21}\rvert}^{2}P_{1}}{N_{0} }\Big) \label{eqn71}, \\
C_{2} \  = \  \log_{2} \Big( 1 + \frac{{\lvert h_{12}\rvert}^{2}P_{2}}{N_{0}}\Big) \label{eqn72}, \\
C_{E} \  = \  \log_{2} \Big( 1 + \frac{ {\lvert z_{1}\rvert}^{2}P_{1} + {\lvert z_{2}\rvert}^{2}P_{2} }{ N_{0} } \Big).
\end{eqnarray}
A secrecy rate pair $(R_{1}, R_{2})$ which falls in the following region is achievable \cite{ic22}:
\begin{eqnarray}
0 \ \leq \ R_{1} \ \leq  \ R^{'}_{1}, \quad
0 \ \leq \ R_{2} \ \leq  \ R^{'}_{2}, \nonumber \\
0 \ \leq \ R_{1} + R_{2} \ \leq \ R^{'}_{1} + R^{'}_{2} - R^{'}_{E}, \nonumber \\
{[P_{1s} + P_{1n}, \ P_{2s} + P_{2n}]}^{T} \ \leq \ {[P_{1}, \ P_{2}]}^{T}, \nonumber \\ {[P_{1s}, \ P_{1n}, \ P_{2s}, \ P_{2n}]}^{T} \ \geq \ {[0, \ 0, \ 0, \ 0]}^{T}.\label{eqn73}
\end{eqnarray}
We intend to maximize the sum secrecy rate subject to the power constraint, 
i.e.,
\begin{eqnarray}
\max_{P_{1s}, \ P_{1n}, \atop{P_{2s}, \ P_{2n}}} \ R^{'}_{1} + R^{'}_{2} - R^{'}_{E} & & \label{eqn74} \\
& \hspace{-60mm} = & \hspace{-30mm} \max_{P_{1s}, \ P_{1n}, \atop{P_{2s}, \ P_{2n}}} \ \Big \{\log_{2} \Big( 1 + \frac{{\lvert h_{21}\rvert}^{2}P_{1s}}{N_{0} + {\lvert h_{21}\rvert}^{2}P_{1n}}\Big) \nonumber \\ 
& \hspace{-50mm} & \hspace{-30mm} +\log_{2} \Big( 1 + \frac{{\lvert h_{12}\rvert}^{2}P_{2s}}{N_{0} + {\lvert h_{12}\rvert}^{2}P_{2n}}\Big) \nonumber \\ 
&\hspace{-50mm} & \hspace{-30mm}- \log_{2} \Big( 1 + \frac{ {\lvert z_{1}\rvert}^{2}P_{1s} + {\lvert z_{2}\rvert}^{2}P_{2s} }{ N_{0} + {\lvert z_{1} \rvert}^{2}P_{1n} + {\lvert z_{2}\rvert}^{2}P_{2n} } \Big) \Big \} \label{eqn75}
\end{eqnarray}
\begin{eqnarray}
\text{s.t.} \quad {[P_{1s} + P_{1n}, \ P_{2s} + P_{2n}]}^{T} \ \leq \ {[P_{1}, \ P_{2}]}^{T}, \nonumber \\ {[P_{1s}, \ P_{1n}, \ P_{2s}, \ P_{2n}]}^{T} \ \geq\ {[0, \ 0, \ 0, \ 0]}^{T}. \label{eqn76}
\end{eqnarray}
This is a nonlinear optimization problem, and we solve it using 
two-dimensional search as follows.

$\textbf{Step1 :}$ Divide the intervals $[0, C_1]$ and $[0, C_2]$ in $K$ and 
$L$ small intervals, respectively, of size $\triangle_{1} = \frac{C_1}{K}$ 
and $\triangle_{2} = \frac{C_2}{L}$ where $K$ and $L$ are large integers.
Let $R^{'k}_{1} = k \triangle_{1}$ and $R^{'l}_{2} = l \triangle_{2}$, 
where $k = 0,1,2,\cdots,K$ and $l = 0,1,2,\cdots,L$.

$\textbf{Step2 :}$ For a given $(R^{'k}_{1}, \ R^{'l}_{2})$ pair, we minimize 
$R^{'}_{E}$ as follows:
\begin{eqnarray}
R^{''kl}_{E} \Define \min_{P_{1s}, \ P_{1n}, \atop{ P_{2s}, \ P_{2n}}} \ \log_{2} \Big( 1 + \frac{ {\lvert z_{1}\rvert}^{2}P_{1s} + {\lvert z_{2}\rvert}^{2}P_{2s} }{ N_{0} + {\lvert z_{1} \rvert}^{2}P_{1n} + {\lvert z_{2}\rvert}^{2}P_{2n} } \Big) \label{eqn77} 
\end{eqnarray}
s.t.
\begin{eqnarray}
R^{''k}_{1} \ \Define \ \log_{2} \Big( 1 + \frac{{\lvert h_{21}\rvert}^{2}P_{1s}}{N_{0} + {\lvert h_{21}\rvert}^{2}P_{1n}}\Big) \ \geq \ R^{'k}_{1}, \nonumber \\
R^{''l}_{2} \ \Define \  \log_{2} \Big( 1 + \frac{{\lvert h_{12}\rvert}^{2}P_{2s}}{N_{0} + {\lvert h_{12}\rvert}^{2}P_{2n}}\Big) \ \geq \ R^{'l}_{2}, \nonumber \\
{[P_{1s} + P_{1n}, \ P_{2s} + P_{2n}]}^{T} \ \leq \ {[P_{1}, \ P_{2}]}^{T}, \nonumber \\ {[P_{1s}, \ P_{1n}, \ P_{2s}, \ P_{2n}]}^{T} \ \geq\ {[0, \ 0, \ 0, \ 0]}^{T}. \label{eqn78}
\end{eqnarray}
The maximum sum secrecy rate is given by $\max_{k = 0,1,2,\cdots,K, \atop{l = 0,1,2,\cdots,L}} \ (R^{''k}_{1} + R^{''l}_{2} - R^{''kl}_{E})$.

We solve the optimization problem (\ref{eqn77}) as follows. Dropping the 
logarithm in the objective function in (\ref{eqn77}), we rewrite the 
optimization problem (\ref{eqn77}) in the following equivalent form:
\begin{eqnarray}
\min_{t, \ P_{1s}, \ P_{1n}, \atop{ P_{2s}, \ P_{2n}}} \ \ t 
\label{eqn79} 
\end{eqnarray}
s.t.
{\small
\begin{eqnarray}
\big( {\lvert z_{1}\rvert}^{2}P_{1s} + {\lvert z_{2}\rvert}^{2}P_{2s} \big) - t \big( N_{0} + {\lvert z_{1} \rvert}^{2}P_{1n} + {\lvert z_{2}\rvert}^{2}P_{2n} \big) \ \leq \ 0, \nonumber \\
\big( 2^{R^{'k}_{1}} - 1 \big) \big( N_{0} + {\lvert h_{21}\rvert}^{2}P_{1n} \big) - \big( {\lvert h_{21}\rvert}^{2}P_{1s} \big) \ \leq \ 0, \nonumber \\
\big( 2^{R^{'l}_{2}} - 1 \big) \big( N_{0} + {\lvert h_{12}\rvert}^{2}P_{2n} \big) - \big( {\lvert h_{12}\rvert}^{2}P_{2s} \big) \ \leq \ 0, \nonumber \\
{[P_{1s} + P_{1n}, \ P_{2s} + P_{2n}]}^{T} \ \leq \ {[P_{1}, \ P_{2}]}^{T}, \nonumber \\ {[P_{1s}, \ P_{1n}, \ P_{2s}, \ P_{2n}]}^{T} \ \geq\ {[0, \ 0, \ 0, \ 0]}^{T}. \label{eqn80}
\end{eqnarray}
}

\vspace{-4mm}
\hspace{-4mm}
For a given $t$, the above problem is formulated as the following linear
feasibility problem \cite{ic16}: 
\begin{eqnarray}
\text{find} \quad P_{1s}, \ P_{1n}, \ P_{2s}, \ P_{2n}, \label{eqn81}
\end{eqnarray}
subject to the constraints in (\ref{eqn80}). 
The minimum value of $t$, denoted by $t^{kl}_{min}$, can be obtained using 
bisection method as follows. Let $t^{kl}_{min}$ lie in the interval 
$[t_{ll}, t_{ul}]$. The value of $t_{ll}$ can be taken as 0 (corresponding 
to the minimum information rate of 0) and $t_{ul}$ can be taken as 
$(2^{C^{}_{E}} - 1)$, which corresponds to the 
information capacity of the eavesdropper link.
Check the feasibility of 
(\ref{eqn81}) at $t^{kl}_{min} = (t^{}_{ll} + t^{}_{ul})/2$. If feasible, 
then $t^{}_{ul} = t^{kl}_{min}$, else $ \ t^{}_{ll} = t^{kl}_{min}$. 
Repeat this until $t^{}_{ul} - t^{}_{ll} \leq \zeta$, where $\zeta$ is 
a small positive number. Using $t^{kl}_{min}$ in (\ref{eqn77}), 
$R^{''kl}_{E}$ is given by
\begin{eqnarray}
R^{''kl}_{E} \ = \ \log_2 (1 + t^{kl}_{min}). \label{eqn82}
\end{eqnarray}

\section{Sum secrecy rate - imperfect CSI}
\label{sec4}
In this section, we assume that the available CSI in all the links are 
imperfect, i.e.,
\begin{eqnarray}
h_{11} = {h}^{0}_{11} + e_{11}, \quad h_{12} = {h}^{0}_{12} + e_{12},  
\label{eqn83} \\
h_{21} = {h}^{0}_{21} + e_{21}, \quad h_{22} = {h}^{0}_{22} + e_{22},  
\label{eqn84} \\
z_{1}  = {z}^{0}_{1}  + e_{1},  \quad z_{2}  = {z}^{0}_{2}  + e_{2},   
\label{eqn85} 
\end{eqnarray}
where ${h}^{0}_{11}$, ${h}^{0}_{12}$, ${h}^{0}_{21}$, ${h}^{0}_{22}$, 
${z}^{0}_{1}$, and  ${z}^{0}_{2}$ are the estimates of ${h}^{}_{11}$, 
${h}^{}_{12}$, ${h}^{}_{21}$, ${h}^{}_{22}$, ${z}^{}_{1}$, and  
${z}^{}_{2}$, respectively, and ${e}^{0}_{11}$, ${e}^{0}_{12}$, 
${e}^{0}_{21}$, ${e}^{0}_{22}$, ${e}^{0}_{1}$, and  ${e}^{0}_{2}$
are the corresponding errors. We assume that errors are bounded in 
their respective absolute values as 
\begin{eqnarray}
{\lvert e_{11} \rvert}^{2} \leq {\epsilon}^{2}_{11}, \quad {\lvert e_{12} \rvert}^{2} \leq {\epsilon}^{2}_{12}, \label{eqn86} \\
{\lvert e_{21} \rvert}^{2} \leq {\epsilon}^{2}_{21}, \quad {\lvert e_{22} \rvert}^{2} \leq {\epsilon}^{2}_{22}, \label{eqn87} \\
{\lvert e_{1} \rvert}^{2} \leq {\epsilon}^{2}_{1}, \quad {\lvert e_{2} \rvert}^{2} \leq {\epsilon}^{2}_{2}. \label{eqn88} 
\end{eqnarray}
With the above error model, we rewrite (\ref{eqn8}), (\ref{eqn9}), and 
(\ref{eqn3}) as follows:
\begin{eqnarray}
y^{'}_{1} \ = \ y_{1} - h^{0}_{11}(X^{1} + N^{1}) \nonumber \\ = \ e^{0}_{11}(X^{1} + N^{1}) + (h^{0}_{12} + e_{12})(X^{2} + N^{2}) + \eta_{1}, \label{eqn89} \\
y^{'}_{2} \ = \ y_{2} - h^{0}_{22}(X^{2} + N^{2}) \nonumber \\ = \ (h^{0}_{21} + e_{21})(X^{1} + N^{1}) + e^{0}_{22}(X^{2} + N^{2}) + \eta_{2}, \label{eqn90} \\
y_{E}      \ = \ (z^{0}_{1} + e_{1})(X^{1} + N^{1}) +  (z^{0}_{2} + e_{2})(X^{2} + N^{2}) \nonumber \\ + \eta_{E}. \label{eqn91} 
\end{eqnarray}
In order to compute $R^{'k}_{1}$, $R^{'l}_{2}$, and $R^{''kl}_{E}$, 
respectively, as described in $\textbf{Step1}$ and $\textbf{Step2}$ in 
Section \ref{sec3}, we get the capacities $C_{1}$, $C_{2}$, and $C_{E}$ 
with imperfect CSI as follows:
\begin{eqnarray}
C_{1} = \min_{e_{21}} \ \log_{2} \Big( 1 + \frac{{\lvert h^{0}_{21} + e_{21}\rvert}^{2}P_{1}}{N_{0}}\Big), \ \text{s.t.} \ {\lvert e_{21} \rvert}^{2} \leq {\epsilon}^{2}_{21},\label{eqn97} \\
= \log_{2} \Big( 1 + \frac{{\lvert \lvert h^{0}_{21}\rvert - \epsilon_{21}\rvert}^{2}P_{1}}{N_{0}}\Big). \label{eqn100}  \\
C_{2} = \min_{e_{12}} \ \log_{2} \Big( 1 + \frac{{\lvert h^{0}_{12} + e_{12} \rvert}^{2}P_{2}}{N_{0}}\Big), \ \text{s.t.} \ {\lvert e_{12} \rvert}^{2} \leq {\epsilon}^{2}_{12},  \label{eqn98} \\
= \log_{2} \Big( 1 + \frac{{\lvert \lvert h^{0}_{12}\rvert - \epsilon_{12}\rvert}^{2}P_{2}}{N_{0}}\Big). \label{eqn101} \\
C_{E} = \max_{e_{1}, \ e_{2}} \  \log_{2} \Big( 1 + \frac{ {\lvert z^{0}_{1} + e_{1} \rvert}^{2}P_{1} + {\lvert z^{0}_{2} + e_{2}\rvert}^{2}P_{2} }{ N_{0} } \Big), \nonumber \\ \text{s.t.} \quad {\lvert e_{1} \rvert}^{2} \leq {\epsilon}^{2}_{1},  \ {\lvert e_{2} \rvert}^{2} \leq {\epsilon}^{2}_{2}, \label{eqn99} \\
= \log_{2} \Big( 1 + \frac{ {\lvert \lvert z^{0}_{1} \rvert +  \epsilon_{1}  \rvert}^{2}P_{1} + {\lvert \lvert z^{0}_{2} \rvert +  \epsilon_{2} \rvert}^{2}P_{2} }{ N_{0} } \Big). \label{102}
\end{eqnarray}
Using (\ref{eqn89}), (\ref{eqn90}), and (\ref{eqn91}), we rewrite the 
optimization problem (\ref{eqn77}) as follows:
\begin{eqnarray}
R^{''kl}_{E}  \Define  \min_{P_{1s}, \ P_{1n}, \atop{P_{2s}, \ P_{2n}}} \ \max_{e_{11}, \ e_{12}, \ e_{21}, \atop{ e_{22}, \ e_{1}, \ e_{2}}} \ \log_{2} \nonumber \\ \Big( 1 +  \frac{ {\lvert (z^{0}_{1} + e_{1})\rvert}^{2}P_{1s} + {\lvert (z^{0}_{2} + e_{2})\rvert}^{2}P_{2s} }{ N_{0} + {\lvert (z^{0}_{1} + e_{1}) \rvert}^{2}P_{1n} + {\lvert (z^{0}_{2} + e_{2})\rvert}^{2}P_{2n} } \Big) \label{eqn92}  
\end{eqnarray}
s.t.
{\small
\begin{eqnarray}
R^{''k}_{1} \Define \min_{e_{21}, \ e_{22}} \ \log_{2} \nonumber \\ \Big( 1 +  \frac{{\lvert h^{0}_{21} + e_{21}\rvert}^{2}P_{1s}}{N_{0} + {\lvert e_{22}\rvert}^{2}(P_{2s} + P_{2n}) + {\lvert h^{0}_{21} + e_{21}\rvert}^{2}P_{1n}}\Big) \ \geq \ R^{'k}_{1}, \label{eqn93} \\
R^{''l}_{2} \Define \min_{e_{11}, \ e_{12}} \ \log_{2} \nonumber \\ \Big( 1 +  \frac{{\lvert h^{0}_{12} + e_{12} \rvert}^{2}P_{2s}}{N_{0} + {\lvert e_{11}\rvert}^{2}(P_{1s} + P_{1n}) + {\lvert h^{0}_{12} + e_{12}\rvert}^{2}P_{2n}}\Big) \ \geq \ R^{'l}_{2}, \label{eqn94} \\
{\lvert e_{11} \rvert}^{2} \leq {\epsilon}^{2}_{11}, \quad {\lvert e_{12} \rvert}^{2} \leq {\epsilon}^{2}_{12}, \quad
{\lvert e_{21} \rvert}^{2} \leq {\epsilon}^{2}_{21}, \nonumber \\
{\lvert e_{22} \rvert}^{2} \leq {\epsilon}^{2}_{22}, \quad 
{\lvert e_{1} \rvert}^{2} \leq {\epsilon}^{2}_{1}, \quad {\lvert e_{2} \rvert}^{2} \leq {\epsilon}^{2}_{2}. \label{eqn95} \\
{[P_{1s} + P_{1n}, \ P_{2s} + P_{2n}]}^{T} \ \leq \ {[P_{1}, \ P_{2}]}^{T}, \nonumber \\ {[P_{1s}, \ P_{1n}, \ P_{2s}, \ P_{2n}]}^{T} \ \geq\ {[0, \ 0, \ 0, \ 0]}^{T}. \label{eqn96}
\end{eqnarray}
}

\vspace{-4mm}
\hspace{-5mm}
In the constraints (\ref{eqn93}) and (\ref{eqn94}), additional noise appear 
due the terms $e^{0}_{22}(X^{2} + N^{2})$ and $e^{0}_{11}(X^{1} + N^{1})$, 
respectively, which have been treated as self noise. For a given 
$(R^{'k}_{1}, \ R^{'l}_{2})$ pair, the worst case sum secrecy rate is 
given by $\max_{k = 0,1,2,\cdots,K, \atop{l = 0,1,2,\cdots,L}} \ (R^{''k}_{1} + R^{''l}_{2} - R^{''kl}_{E})$.

We solve the optimization problem (\ref{eqn92}) as follows.
We write the optimization problem (\ref{eqn92}) in the following form:
\begin{eqnarray}
\min_{P_{1s}, \ P_{1n}, \atop{P_{2s}, \ P_{2n}}} \ \max_{e_{11}, \ e_{12}, \ e_{21}, \atop{ e_{22}, \ e_{1}, \ e_{2}}} \nonumber \\ \Big( \frac{ {\lvert (z^{0}_{1} + e_{1})\rvert}^{2}P_{1s} + {\lvert (z^{0}_{2} + e_{2})\rvert}^{2}P_{2s} }{ N_{0} + {\lvert (z^{0}_{1} + e_{1}) \rvert}^{2}P_{1n} + {\lvert (z^{0}_{2} + e_{2})\rvert}^{2}P_{2n} } \Big) \label{eqn102}  
\end{eqnarray}
s.t.
\begin{eqnarray}
\min_{e_{21}, \ e_{22}} \Big( \frac{{\lvert h^{0}_{21} + e_{21}\rvert}^{2}P_{1s}}{N_{0} + {\lvert e_{22}\rvert}^{2}(P_{2s} + P_{2n}) + {\lvert h^{0}_{21} + e_{21}\rvert}^{2}P_{1n}}\Big) \nonumber \\ \geq \ (2^{R^{'k}_{1}} -1),  \nonumber \\
\min_{e_{11}, \ e_{12}} \Big( \frac{{\lvert h^{0}_{12} + e_{12} \rvert}^{2}P_{2s}}{N_{0} + {\lvert e_{11}\rvert}^{2}(P_{1s} + P_{1n}) + {\lvert h^{0}_{12} + e_{12}\rvert}^{2}P_{2n}}\Big) \nonumber \\ \geq \ (2^{R^{'l}_{2}} -1),  \nonumber \\
{\lvert e_{11} \rvert}^{2} \leq {\epsilon}^{2}_{11}, \quad {\lvert e_{12} \rvert}^{2} \leq {\epsilon}^{2}_{12}, \quad
{\lvert e_{21} \rvert}^{2} \leq {\epsilon}^{2}_{21}, \nonumber \\
 {\lvert e_{22} \rvert}^{2} \leq {\epsilon}^{2}_{22}, \quad
{\lvert e_{1} \rvert}^{2} \leq {\epsilon}^{2}_{1}, \quad {\lvert e_{2} \rvert}^{2} \leq {\epsilon}^{2}_{2}.  \nonumber \\
{[P_{1s} + P_{1n}, \ P_{2s} + P_{2n}]}^{T} \ \leq \ {[P_{1}, \ P_{2}]}^{T}, \nonumber \\ {[P_{1s}, \ P_{1n}, \ P_{2s}, \ P_{2n}]}^{T} \ \geq\ {[0, \ 0, \ 0, \ 0]}^{T}. \label{eqn103}
\end{eqnarray}
We get the following upper bound for the above optimization problem:
\begin{eqnarray}
\min_{P_{1s}, \ P_{1n}, \atop{P_{2s}, \ P_{2n}}} \ \min_{e_{12}, \ e_{21}, \ e_{1}, \ e_{2}, \atop{t_{1}, \ t_{2}, \ t_{3}, \ t_{4}, \atop{t_{5}, \ t_{6}, \ t_{7}, \ t_{8}}}} \ \Big( \frac{t_{1} + t_{2}}{N_{0} + t_{3} + t_{4}} \Big) \label{eqn104} 
\end{eqnarray}
\begin{eqnarray}
\text{s.t.} \quad t_{3} \ \geq \ 0, \ t_{4} \ \geq \ 0, \ t_{5} \ \geq \ 0, \ t_{7} \ \geq \ 0, \label{eqn110} \\
\forall e_{1} \quad \text{s.t.} \quad {\lvert e_{1} \rvert}^{2} \leq {\epsilon}^{2}_{1} \ \Longrightarrow \ \nonumber \\  {\lvert (z^{0}_{1} + e_{1}) \rvert }^{2}P_{1s} - t_{1} \ \leq \ 0, \label{eqn111} \\
\forall e_{1} \quad \text{s.t.} \quad {\lvert e_{1} \rvert}^{2} \leq {\epsilon}^{2}_{1} \ \Longrightarrow \ \nonumber \\ -{\lvert (z^{0}_{1} + e_{1}) \rvert }^{2}P_{1n} + t_{3} \ \leq \ 0,  \label{eqn112} \\
\forall e_{2} \quad \text{s.t.} \quad {\lvert e_{2} \rvert}^{2} \leq {\epsilon}^{2}_{2} \ \Longrightarrow \ \nonumber \\  {\lvert (z^{0}_{2} + e_{2}) \rvert }^{2}P_{2s} - t_{2} \ \leq \ 0, \label{eqn113} \\
\forall e_{2} \quad \text{s.t.} \quad {\lvert e_{2} \rvert}^{2} \leq {\epsilon}^{2}_{2} \ \Longrightarrow \ \nonumber \\ -{\lvert (z^{0}_{2} + e_{2}) \rvert }^{2}P_{2n} + t_{4} \ \leq \ 0, \label{eqn114} \\
\Big( \frac{t_{5}}{N_{0} + {\lvert \epsilon_{22}\rvert}^{2}(P_{2s} + P_{2n}) + t_{6}}\Big) \ \geq \ (2^{R^{'k}_{1}} -1),  \label{eqn115} \\
\forall e_{21} \quad \text{s.t.} \quad {\lvert e_{21} \rvert}^{2} \leq {\epsilon}^{2}_{21} \ \Longrightarrow \ \nonumber \\  -{\lvert (h^{0}_{21} + e_{21}) \rvert }^{2}P_{1s} + t_{5} \ \leq \ 0 ,  \label{eqn116} \\
\forall e_{21} \quad \text{s.t.} \quad {\lvert e_{21} \rvert}^{2} \leq {\epsilon}^{2}_{21} \ \Longrightarrow \ \nonumber \\ {\lvert (h^{0}_{21} + e_{21}) \rvert }^{2}P_{1n} - t_{6} \ \leq \ 0, \label{eqn117} \\
\Big( \frac{t_{7}}{N_{0} + {\lvert \epsilon_{11}\rvert}^{2}(P_{1s} + P_{1n}) + t_{8}}\Big) \ \geq \ (2^{R^{'l}_{2}} -1),  \label{eqn118} \\
\forall e_{12} \quad \text{s.t.} \quad {\lvert e_{12} \rvert}^{2} \leq {\epsilon}^{2}_{12} \ \Longrightarrow \ \nonumber \\ -{\lvert (h^{0}_{12} + e_{12}) \rvert }^{2}P_{2s} + t_{7} \ \leq \ 0 ,  \label{eqn119} \\
\forall e_{12} \quad \text{s.t.} \quad {\lvert e_{12} \rvert}^{2} \leq {\epsilon}^{2}_{12} \ \Longrightarrow \ \nonumber \\ {\lvert (h^{0}_{12} + e_{12}) \rvert }^{2}P_{2n} - t_{8} \ \leq \ 0, \label{eqn120} \\
{[P_{1s} + P_{1n}, \ P_{2s} + P_{2n}]}^{T} \ \leq \ {[P_{1}, \ P_{2}]}^{T}, \label{eqn121} \\ 
{[P_{1s}, \ P_{1n}, \ P_{2s}, \ P_{2n}]}^{T} \ \geq\ {[0, \ 0, \ 0, \ 0]}^{T}. \label{eqn105}
\end{eqnarray}
We use the S-procedure to transform the pairs of quadratic inequalities 
in (\ref{eqn111}), (\ref{eqn112}), (\ref{eqn113}), (\ref{eqn114}), 
(\ref{eqn116}), (\ref{eqn117}), (\ref{eqn119}), and (\ref{eqn120}) to 
equivalent linear matrix inequalities (LMI) \cite{ic16}. We get the 
following single minimization form for the above optimization problem:
\begin{eqnarray}
\min_{P_{1s}, \ P_{1n}, \ P_{2s}, \ P_{2n}, \atop{t_{1}, \ t_{2},\cdots,t_{8}, \atop{\lambda_{1}, \ \lambda_{2},\cdots,\lambda_{8}, \atop{t}}} } \ \ t \label{eqn106} 
\end{eqnarray}
\begin{eqnarray}
\text{s.t.} \quad t_{3} \ \geq \ 0, \ t_{4} \ \geq \ 0, \ t_{5} \ \geq \ 0, \ t_{7} \ \geq \ 0, \nonumber \\
\big( t_{1} + t_{2} \big) - t \big( N_{0} + t_{3} + t_{4}\big) \ \leq \ 0, \nonumber \\
(2^{R^{'k}_{1}} -1) \big( N_{0} + {\lvert \epsilon_{22}\rvert}^{2}(P_{2s} + P_{2n}) + t_{6} \big) -  t_{5} \ \leq \ 0, \nonumber \\
(2^{R^{'l}_{2}} -1) \big( N_{0} + {\lvert \epsilon_{11}\rvert}^{2}(P_{1s} + P_{1n}) + t_{8} \big) -  t_{7}  \ \leq \ 0, \nonumber \\
\left[\footnotesize
\begin{array}{cc}
-P_{1s} + \lambda_{1} & -z^{0}_{1}P_{1s} \\ -z^{0\ast}_{1}P_{1s} & -{\lvert z^{0}_{1}\rvert}^{2}P_{1s} + t_{1} - \lambda_{1}\epsilon^{2}_{1}
\end{array} \right] \succeq \boldsymbol{0}, \quad \lambda_{1} \geq 0, \nonumber \\ 
\left[\footnotesize
\begin{array}{cc}
P_{1n} + \lambda_{2} & z^{0}_{1}P_{1n} \\ z^{0\ast}_{1}P_{1n} & {\lvert z^{0}_{1}\rvert}^{2}P_{1n} - t_{3} - \lambda_{2}\epsilon^{2}_{1}
\end{array} \right] \succeq \boldsymbol{0}, \quad \lambda_{2} \geq 0,  \nonumber \\ 
\left[\footnotesize
\begin{array}{cc}
-P_{2s} + \lambda_{3} & -z^{0}_{2}P_{2s} \\ -z^{0\ast}_{2}P_{2s} & -{\lvert z^{0}_{2}\rvert}^{2}P_{2s} + t_{2} - \lambda_{3}\epsilon^{2}_{2}
\end{array} \right] \succeq \boldsymbol{0}, \quad \lambda_{3} \geq 0, \nonumber \\ 
\left[\footnotesize
\begin{array}{cc}
P_{2n} + \lambda_{4} & z^{0}_{2}P_{2n} \\ z^{0\ast}_{2}P_{2n} & {\lvert z^{0}_{2}\rvert}^{2}P_{2n} - t_{4} - \lambda_{4}\epsilon^{2}_{2}
\end{array} \right] \succeq \boldsymbol{0}, \quad \lambda_{4} \geq 0,  \nonumber \\ 
\left[\footnotesize
\begin{array}{cc}
P_{1s} + \lambda_{5} & h^{0}_{21}P_{1s} \\ h^{0\ast}_{21}P_{1s} & {\lvert h^{0}_{21}\rvert}^{2}P_{1s} - t_{5} - \lambda_{5}\epsilon^{2}_{21}
\end{array} \right] \succeq \boldsymbol{0}, \quad \lambda_{5} \geq 0,  \nonumber \\ 
\left[\footnotesize
\begin{array}{cc}
-P_{1n} + \lambda_{6} & -h^{0}_{21}P_{1n} \\ -h^{0\ast}_{21}P_{1n} & -{\lvert h^{0}_{21}\rvert}^{2}P_{1n} + t_{6} - \lambda_{6}\epsilon^{2}_{21}
\end{array} \right] \succeq \boldsymbol{0}, \quad \lambda_{6} \geq 0, \nonumber \\ 
\left[\footnotesize
\begin{array}{cc}
P_{2s} + \lambda_{7} & h^{0}_{12}P_{2s} \\ h^{0\ast}_{12}P_{2s} & {\lvert h^{0}_{12}\rvert}^{2}P_{2s} - t_{7} - \lambda_{7}\epsilon^{2}_{12}
\end{array} \right] \succeq \boldsymbol{0}, \quad \lambda_{7} \geq 0, \nonumber \\ 
\left[\footnotesize
\begin{array}{cc}
-P_{2n} + \lambda_{8} & -h^{0}_{12}P_{2n} \\ -h^{0\ast}_{12}P_{2n} & -{\lvert h^{0}_{12}\rvert}^{2}P_{2n} + t_{8} - \lambda_{8}\epsilon^{2}_{12}
\end{array} \right] \succeq \boldsymbol{0}, \quad \lambda_{8} \geq 0, \nonumber \\ 
{[P_{1s} + P_{1n}, \ P_{2s} + P_{2n}]}^{T} \ \leq \ {[P_{1}, \ P_{2}]}^{T}, \nonumber \\
{[P_{1s}, \ P_{1n}, \ P_{2s}, \ P_{2n}]}^{T} \ \geq\ {[0, \ 0, \ 0, \ 0]}^{T}. \label{eqn107}
\end{eqnarray}
For a given $t$, the above problem is formulated as the following 
semidefinite feasibility problem \cite{ic16}: 
\begin{eqnarray}
\text{find} \quad P_{1s}, \ P_{1n}, \ P_{2s}, \ P_{2n}, \ t_{1}, \ t_{2},\cdots,t_{8}, \nonumber \\ \lambda_{1}, \ \lambda_{2},\cdots,\lambda_{8},  \label{eqn108}
\end{eqnarray}
subject to the constraints in (\ref{eqn107}). 
The minimum value of $t$, denoted by $t^{kl}_{min}$, can be obtained using 
bisection method as described in section \ref{sec3}. The value of $t_{ll}$ 
can be taken as 0 (corresponding to the minimum information rate of 0). 
The value of $t_{ul}$ can be taken as $(2^{C^{}_{E}} - 1)$, which 
corresponds to the best case information capacity of the eavesdropper link.
Using $t^{kl}_{min}$ in (\ref{eqn92}), $R^{''kl}_{E}$ is given by
\begin{eqnarray}
R^{''kl}_{E} \ = \ \log_2 (1 + t^{kl}_{min}). \label{eqn109}
\end{eqnarray}

\section{Results and Discussions}
\label{sec5}
\begin{figure}[htb]
\includegraphics[totalheight=7.5cm,width=8.5cm]{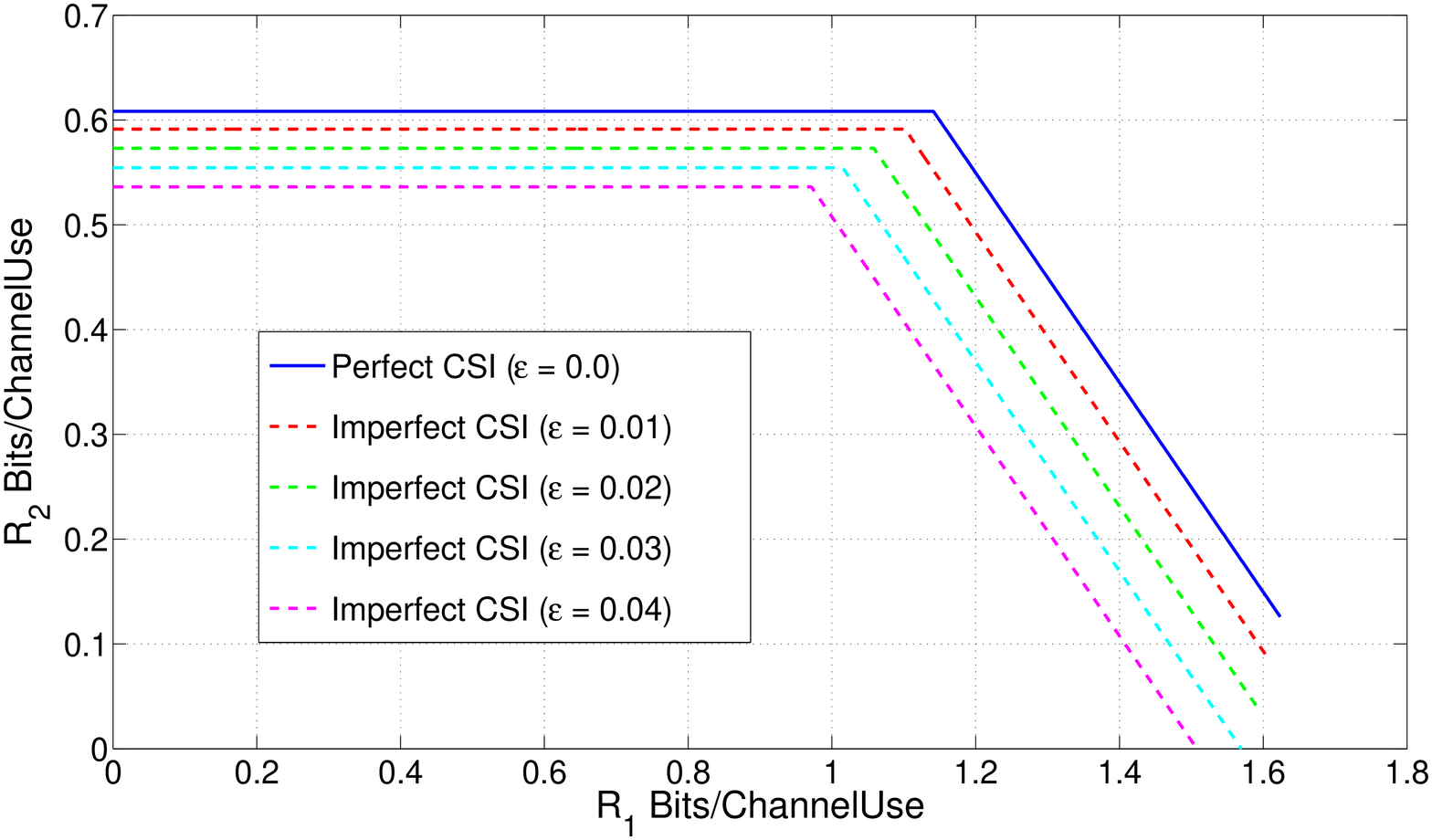}
\caption{Achievable $(R_{1},R_{2})$ region in full-duplex communication. 
$P_{1} = P_{2} = 3$ dB, $\epsilon = 0.0, \ 0.01, \ 0.02, \ 0.03, \ 0.04$, 
and $N_{0} = 1$.}
\label{fig2}
\end{figure}

\begin{figure}[htb]
\includegraphics[totalheight=7.5cm,width=8.5cm]{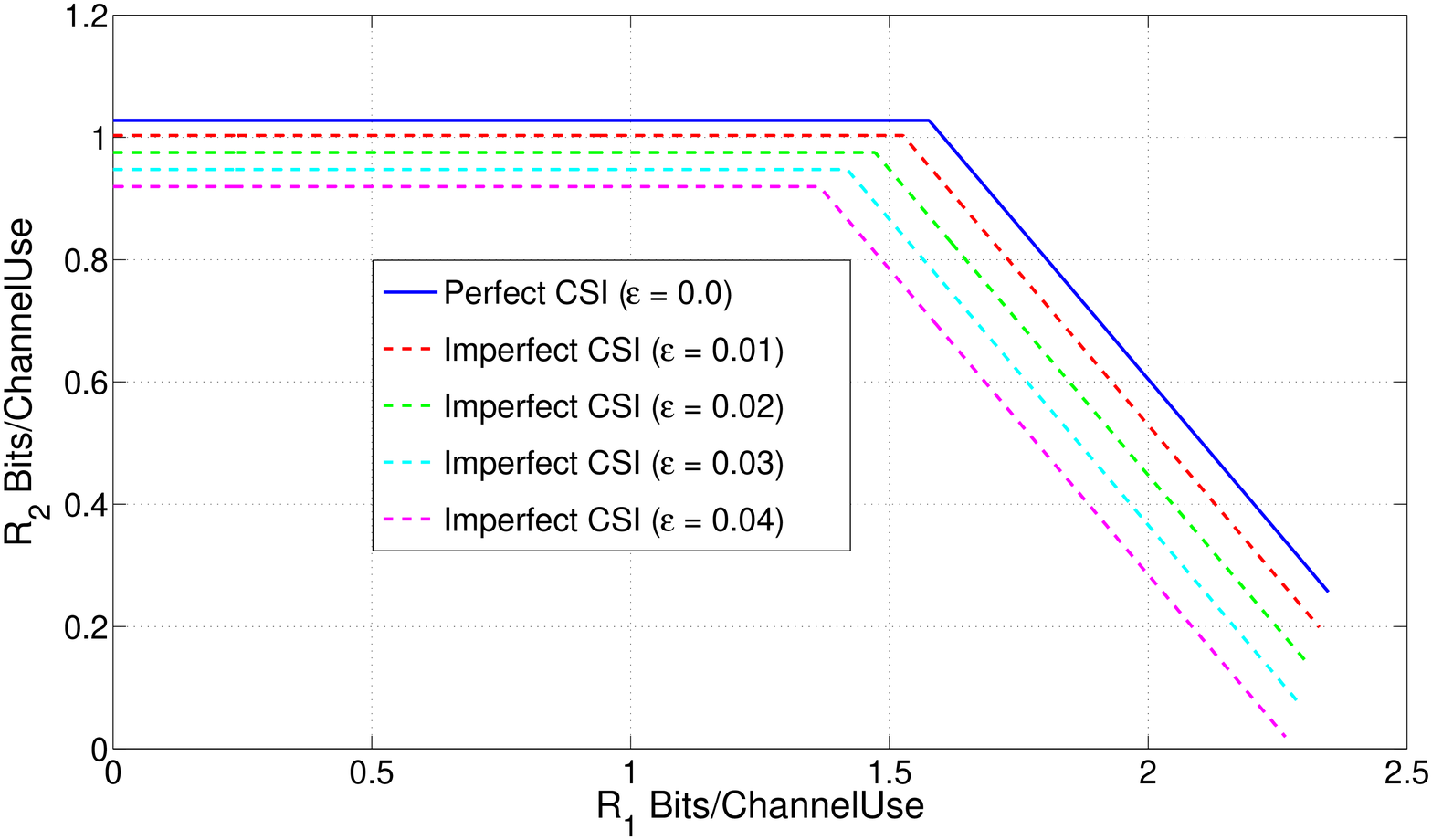}
\caption{Achievable $(R_{1},R_{2})$ region in full-duplex communication. 
$P_{1}= P_{2}=6$ dB, $\epsilon = 0.0, \ 0.01, \ 0.02, \ 0.03, \ 0.04$, 
and $N_{0} = 1$.}
\label{fig3}
\end{figure}

In this section, we present numerical results on the secrecy rate
under perfect and imperfect CSI conditions. We have used the following 
channel gains as the estimates:
$h^{0}_{12} = 0.5054 - 0.1449i$,
$h^{0}_{21} = -0.0878 + 1.0534i$,
$z^{0}_{1} = 0.1187 - 0.2135i$,
$z^{0}_{2} = 0.1268 + 0.2882i$.
We assume that the magnitudes of the CSI errors in all the links are 
equal, i.e., $\epsilon_{11} = \epsilon_{12} = \epsilon_{21} = \epsilon_{22} = \epsilon_{1} = \epsilon_{2} = \epsilon$.
We also assume that $N_{0} = 1$. In Fig. \ref{fig2} and Fig. \ref{fig3},
we plot the $(R_{1},R_{2})$ region obtained by maximizing the sum secrecy 
rate for various values of $\epsilon = 0.0, \ 0.01, \ 0.02, \ 0.03, \ 0.04$.
Results in Fig. \ref{fig2} and Fig. \ref{fig3} are generated for fixed 
powers $P_{1} = P_{2} = 3$ dB and $P_{1} = P_{2} = 6$ dB, respectively.
We observe that as the magnitude of the CSI error increases the corresponding 
sum secrecy rate decreases which results in the shrinking of the achievable 
rate region. Also, as the power is increased from 3 dB to 6 dB, 
the achievable secrecy rate region increases.

\section{Conclusions}
\label{sec6}
We investigated the sum secrecy rate and the corresponding achievable 
secrecy rate region in full-duplex wiretap channel when the CSI in all 
the links are imperfect. We obtained the optimum powers of the message 
and jamming signals which maximize the sum secrecy rate. Numerical 
results illustrated the impact of imperfect CSI on the achievable 
secrecy rate region.


\begin{thebibliography}{99}
\bibitem{ic1}
A. Wyner, ``The wire-tap channel,'' Bell. Syst Tech. J, vol. 54, no. 8,
pp. 1355-1387, Jan. 1975.

\bibitem{ic2}
I. Csiszar and J. Korner, ``Broadcast channels with confidential messages,'' 
{\em IEEE Trans. Inform. Theory}, vol. IT-24, pp. 339-348, May 1978.

\bibitem{ic3}
S. K. Leung-Yan-Cheong and M. E. Hellman, ``The Gaussian wire-tap channel,'' 
{\em IEEE Trans. Inform. Theory}, vol. IT-24, pp. 451-456, Jul. 1978.

\bibitem{ic4}
Y. Liang, H. V. Poor, and S. Shamai (Shitz), ``Information theoretic 
security,'' {\em Foundations and Trends in Communications and Information 
Theory}, NOW Publishers, vol. 5, no. 4-5, 2009.

\bibitem{ic6}
P. K. Gopala, L. Lai, and H. E. Gamal, ``On the secrecy capacity of fading
channels,'' {\em IEEE Trans. Inform. Theory}, vol. 54, no. 10, pp. 4687-4698,
Oct. 2008.

\bibitem{ic7}
S. Shafiee and S. Ulukus, ``Achievable rates in Gaussian MISO channels with
secrecy constraint,'' {\em Proc. IEEE ISIT'2007}, June 2007.

\bibitem{ic8}
A. Khisti and G. Wornell, ``Secure transmission with multiple antennas-I: The
MISOME wiretap channel,'' {\em IEEE Trans. Inform. Theory}, vol. 56, no. 7,
pp. 3088-3104, Jul. 2010.

\bibitem{ic9}
F. Oggier and B. Hassibi, ``The secrecy capacity of the MIMO wiretap 
channel,'' {\em Proc. IEEE ISIT'2008}, July 2008.

\bibitem{ic10}
A. Khisti and G. Wornell, ``Secure transmission with multiple antennas-II: 
The MIMOME wiretap channel,'' {\em IEEE Trans. Inform. Theory}, vol. 56, 
no. 7, pp. 3088-3104, Jul. 2010.

\bibitem{fd_rice}
A. Sabharwal, P. Schniter, D. Guo, D. W. Bliss, S. Rangarajan, and 
R. Wichman, ``In-band full-duplex wireless: challenges and opportunities,''
arXiv:1311.0456v1 [cs.IT] 3 Nov 2013.

\bibitem{ic11}
M. Duarte and A. Sabharwal, ``Full-duplex wireless communications using 
off-the-shelf radios: feasibility and first results,'' {\em Conference 
Record of the Forty Fourth Asilomar Conference on Signals, Systems and 
Computers (ASILOMAR),} pp. 1558-1562, Nov. 2010.

\bibitem{ic12}
T. Riihonen, S. Werner, and R. Wichman, ``Hybrid full-duplex/half-duplex 
relaying with transmit power adaptation,'' {\em IEEE Trans. Wireless 
Commun.}, vol. 10, no. 9, pp. 3074-3085, Sep. 2011.  

\bibitem{ic13}
T. Riihonen, S. Werner, and R. Wichman, ``Mitigation of loopback 
self-interference in full-duplex MIMO relays,'' {\em IEEE Trans. Signal 
Proc.}, vol. 59, no. 12, pp. 5983-5993, Dec. 2011.  

\bibitem{ic14}
A. Thangaraj, R. K. Ganti and S. Bhashyam, ``Self-interference cancellation 
models for full-duplex wireless communications,'' {\em Proc. SPCOM'2012}, 
July 2012.

\bibitem{ic20}
E. Tekin and A. Yener, ``The general Gaussian multiple-access and two-way 
wiretap channels: achievable rates and cooperative jamming,'' {\em IEEE 
Trans. Inform. Theory}, vol. 54, no. 6, pp. 2735-2751, Jun. 2008.

\bibitem{ic21}
E. Tekin and A. Yener, ``Correction to: ``The Gaussian multiple-access
wire-tap channel'' and ``The general Gaussian multiple-access and two-way 
wiretap channels: achievable rates and cooperative jamming'','' {\em IEEE 
Trans. Inform. Theory}, vol. 56, no. 9, pp. 4762-4763, Sep. 2010.

\bibitem{ic22}
A. E. Gamal, O. O. Koyluoglu, M. Youssef, and H. E. Gamal, ``Achievable 
secrecy rate regions for the two-way wiretap channel,'' {\em IEEE Trans. 
Inform. Theory}, vol. 59, no. 12, pp. 8099-8114, Dec. 2013.

\bibitem{ic16}
S. Boyd and L. Vandenberghe, {\em Convex optimization}, Cambridge Univ. 
Press, 2004.

\end{thebibliography}
\end{document}